%% file: main_icml.tex
\documentclass{article}

\usepackage{microtype}
\usepackage{graphicx}
\usepackage{subcaption}
\usepackage{booktabs, multirow, supertabular} 

\usepackage{hyperref}

\usepackage[accepted]{styles/icml2025/icml2025}

\usepackage{amsmath, amsfonts, amssymb, mathtools, amsthm}

\usepackage[capitalize,noabbrev]{cleveref}

\theoremstyle{plain}
\newtheorem{theorem}{Theorem}[section]

\newtheorem{lemma}[theorem]{Lemma}
\newtheorem{corollary}[theorem]{Corollary}
\theoremstyle{definition}
\newtheorem{definition}[theorem]{Definition}

\newtheorem{example}[theorem]{Example}
\theoremstyle{remark}
\newtheorem{remark}[theorem]{Remark}

\usepackage[textsize=tiny]{todonotes}

\input{texs/newcommands}

\begin{document}

\twocolumn[
\icmltitle{Differentially private synthesis of Spatial Point Processes}



\icmlsetsymbol{equal}{*}

\begin{icmlauthorlist}
\icmlauthor{Dangchan Kim}{yyy,pol}
\icmlauthor{Chae Young Lim}{yyy,idis}
\end{icmlauthorlist}

\icmlaffiliation{yyy}{Department of Statistics, Seoul National University, Seoul, Korea}
\icmlaffiliation{pol}{Korea National Police Agency, Seoul, Korea}
\icmlaffiliation{idis}{Institute for Data Innovation in Science, Seoul National University, Seoul, Korea}

\icmlcorrespondingauthor{Dangchan Kim}{dang112@snu.ac.kr}
\icmlcorrespondingauthor{Chae Young Lim}{twinwood@snu.ac.kr}

\icmlkeywords{Differential Privacy, Data Synthesis, Spatial Point Process}

\vskip 0.3in
]



\printAffiliationsAndNotice{}  

\begin{abstract}
\input{texs/abstract}
\end{abstract}

\section{Introduction}
\input{texs/introduction}

\section{Preliminaries}
\input{texs/preliminaries}

\section{Methodology}
\input{texs/method}

\section{Experiments}
\input{texs/experiments}

\section{Conclusion}
\input{texs/conclusion}

 \section*{Acknowledgements}
 This work was partly supported by the National Research Foundation of Korea (NRF) grant funded by the Korea government (MSIT) (No. RS-2024-00335033) and Institute of Information \& communications Technology Planning \& Evaluation (IITP) grant funded by the Korea government(MSIT) (No.2022-0-00937). 



%


\bibliography{bibliography.bib}
\bibliographystyle{styles/icml2025/icml2025.bst}

\newpage
\appendix
\onecolumn
\input{texs/appendicies}

\end{document}

%% file: texs/newcommands.tex
\newcommand{\lap}{\mathcal{M}_{\rm Lap}}
\newcommand{\lgcp}{\mathcal{M}_{\rm LGCP}}
\newcommand{\kernel}{\mathcal{M}_{\rm Kern}}

\newcommand{\BETA}{\boldsymbol{\beta}}

\newcommand{\GAMMA}{\boldsymbol{\gamma}}
\newcommand{\THETA}{\boldsymbol{\theta}}

\newcommand{\TAU}{\boldsymbol{\tau}}

\newcommand{\x}{\mathbf{x}}
\newcommand{\y}{\mathbf{y}}
\newcommand{\w}{\mathbf{w}}

\newcommand{\X}{\mathbf{X}} 

\newcommand{\cX}{\mathcal{X}}

\newcommand{\N}{\mathbb{N}}

\newcommand{\E}{{\mathrm{E}}}

\newcommand{\cov}{{\rm Cov}}
\newcommand{\var}{{\rm Var}}

\newcommand{\R}{{\mathbb{R}}}

\newcommand{\M}{{\mathcal{M}}}
\newcommand{\ed}{{(\epsilon,\delta)}}

\newcommand{\Po}{\mathrm{Poisson}}
\newcommand{\lf}{\mathrm{lf}}
\newcommand{\nb}{\overset{\alpha}{\sim}}


%% file: texs/abstract.tex
This paper proposes a method to generate synthetic data for spatial point patterns within the differential privacy (DP) framework. Specifically, we define a differentially private Poisson point synthesizer (PPS) and Cox point synthesizer (CPS) to generate synthetic point patterns with the concept of the $\alpha$-neighborhood that relaxes the original definition of DP. We present three example models to construct a differentially private PPS and CPS, providing sufficient conditions on their parameters to ensure the DP given a specified privacy budget. In addition, we demonstrate that the synthesizers can be applied to point patterns on the linear network. Simulation experiments demonstrate that the proposed approaches effectively maintain the privacy and utility of synthetic data. 

%% file: texs/introduction.tex
Preserving privacy has become increasingly important with the growing collection and analysis of data. As data availability grows, so does the risk of misuse or unintended disclosure of sensitive information. Differential privacy (DP), introduced by \citet{dworkDifferentialPrivacy2006}, has emerged as the de facto standard framework for protecting individual privacy. DP provides privacy guarantees by modifying data analysis processes to limit the exposure of personal data probabilistically. This approach allows for data analysis while maintaining privacy through the use of differentially private machine learning models or the release of private estimators.

In certain cases, releasing datasets to the public becomes necessary. Data anonymization is a common choice to release a dataset without personal information \citep{bayardoDataPrivacyOptimal2005}, but it does not guarantee privacy for further analysis. The framework of differential privacy is subsequently extended to the data release mechanism, a function generating synthetic data from the original dataset. Deep generative models such as generative adversarial networks (GANs) are widely used for generating differentially private synthetic data \citep{jordonPATEGANGENERATINGSYNTHETIC2019}. Additionally, certain statistical models can also serve this purpose \citep{liuModelbasedDifferentiallyPrivate2016}.

A spatial point pattern is a dataset in which each data point represents a location of a random event. It is widely used in various fields such as geology, epidemiology, and crime research \citep{baddeleySpatialPointPatterns2016}. These datasets pose significant privacy risks, as they can potentially reveal sensitive information such as individuals' residential locations. Various methods have been proposed to generate synthetic spatial point patterns in a differentially private manner \citep{qardajiDifferentiallyPrivateGrids2013, shahamHTFHomogeneousTree2021, cunninghamGeoPointGANSyntheticSpatial2022, ahujaNeuralApproachSpatioTemporal2023}. While these approaches including differentially private histograms and generative models work well in many cases, they might struggle with structured spatial domains like linear networks. In such domains, spatial relationships are more complex than in traditional Euclidean spaces \citep{baddeleyAnalysingPointPatterns2021a}.

A statistical model-based approach can provide privacy protection in such scenarios by utilizing both the data's statistical properties and its spatial domain. Previous studies have explored model-based synthesis of spatial point patterns using Cox processes \citep{quickBayesianMarkedPoint2015,walderPrivacySpatialPoint2020}. However, these methods ensure privacy only in terms of disclosure risk, not differential privacy. The key challenge in developing model-based differentially private synthesis for spatial point patterns lies in the strict constraints imposed on the intensity function of the underlying point process.

To address the inutilizability of model-based point synthesis, we propose an approach that relaxes the definition of DP within the spatial point pattern context. Specifically, we introduce a differential privacy under neighborhood that defines neighboring datasets under bounded perturbation. Our main contribution is differentially private point synthesizers, the \textit{Poisson point synthesizer} and the \textit{Cox point synthesizer} derived from the Poisson point processes and the Cox processes, respectively. Then, we provide three concrete example classes: (1) kernel intensity estimation (2) log Gaussian Cox processes, and (3) the Laplace mechanism. We also establish theoretical guarantees under which each proposed approach ensures differential privacy. Moreover, we extend the Cox point synthesizers to spatial point processes defined on linear networks with resistance metric-based covariance functions proposed by \citet{anderesIsotropicCovarianceFunctions2020}. We also provide a simulation study to evaluate the utility of the synthetic data generated by the Poisson and the Cox point synthesizers. Additionally, we apply our method to a real-world linear network case study using the Chicago crime data.

%% file: texs/preliminaries.tex
In this section, we provide a brief introduction to differential privacy and spatial point processes.

\subsection{Differential privacy}
\label{sec:dp}

A dataset $D$ is a collection of $n$ random samples $D=(X_1, \cdots,X_n) \in\cX^n$ where $\cX$ is the space of all possible records. Two datasets $D_1, D_2 \in \cX^n$ are \textit{neighboring} if they differ in only one record, and we denote this as $D_1\sim D_2$. Given dataset $D$, a \textit{randomized mechanism} $\M:\cX^n\to F$ is a function that maps $D$ to a value in the output space $F$ endowed with its $\sigma$-field $\mathcal{F}$.

\begin{definition}[\citeauthor{dworkDifferentialPrivacy2006}]
     A randomized mechanism $\M:\cX^n\to F$ is said to achieve \textit{$(\epsilon,\delta)$-differential privacy} if for any measurable set $A\in\mathcal{F}$ and any neighboring $D_1$ and $D_2$,
    \begin{equation}
        P (\M (D_1)\in A) \leq e^\epsilon P (\M (D_2)\in A) + \delta.
    \end{equation}
\end{definition}

The $(\epsilon, \delta)$-differential privacy is called approximate DP, while $(\epsilon, 0)$-differential privacy is known as pure DP. The parameter $\epsilon$, or \textit{privacy budget} represents privacy strength, with smaller $\epsilon$ indicating stronger protection.

A common approach for achieving differential privacy involves adding independently and identically distributed (i.i.d.) noise, such as noise following a Laplace distribution to the output of a function (see \cref{ex:laplace}) \citep{dworkAlgorithmicFoundationsDifferential2013}.

\begin{example}[Laplace mechanism]
Randomized mechanism outputting $\M (D) = \tilde f (D)$ where
\begin{equation}
\tilde f (D) = f (D) + \text{Lap}\left (\dfrac{\Delta_1(f)}{\epsilon}\right)
\end{equation}
achieves $(\epsilon,0)$-differential privacy. $\text{Lap}(\lambda)$ is an i.i.d. Laplace random vector with mean zero and scale $\lambda$. $\Delta_1(f) = \max_{D_1\sim D_2}\Vert f (D_1)-f (D_2)\Vert_1$ is the \textit{L1 sensitivity} of $f$.
\label{ex:laplace}
\end{example}

\subsection{Spatial point process}
\label{sec:point_process}

A spatial point pattern is a dataset that records the observed spatial locations of events. A spatial point process is a statistical model for the spatial point pattern, which is a countable random subset of a space $S$. Throughout this paper, we consider the spatial domain $S$ as a compact subset of $\R^2$ and refer to $S$ as the \textit{spatial domain}. Following \citet{mollerStatisticalInferenceSimulation2003}, we consider the spatial point processes $\X$ of which realizations are locally finite. That is, a realization of $\X$ takes values in the set of locally finite configurations $N_\lf = \{\x \subseteq S: n (\x_B) < \infty \text{ for all bounded } B\subseteq S\}$, where $n (\x_B)$ denotes the number of points in the bounded region $B$. Also, let $\mathcal{N}_\lf$ be the $\sigma$-field generated by $N_\lf$.

A Poisson point process is a simple and widely used model for the spatial point process, where the number of points falling in a region is given by the Poisson random variable and the points are independently distributed over the spatial domain by the governed intensity function given the number of points.

The Poisson point process $\X$ is uniquely determined by its intensity measure $\mu$, which is given by $\mu(A) = \E[N(A)]$ for any $A\in\mathcal{S}$, where $N (A)$ is the number of points falling in $A$. The intensity function $\lambda$ is related to $\mu$ by $\mu (B) = \int_B \lambda (s)ds$ for any $B\subseteq S$. Also, the intensity measure $\mu$ is locally finite in the sense that $\mu (B)<\infty$ for any bounded $B\subseteq S$ \citep{mollerStatisticalInferenceSimulation2003}.

A Cox process extends the Poisson point process by allowing the intensity function to be random. Specifically, the conditional distribution of a Cox process given its intensity function is the Poisson point process. In this paper, we use the log-Gaussian Cox process (LGCP), where the random intensity function is given by the exponential of a Gaussian process \citep{mollerLogGaussianCox1998}. The definitions of the Poisson point process, log-Gaussian Cox process, and their counterparts on a linear network are given in \cref{app:pointprocess}.

%% file: texs/method.tex
In this section, we introduce differential privacy under the $\alpha$-neighborhood framework for spatial point patterns. We then define point synthesizers with three example approaches: kernel intensity estimation, the log-Gaussian Cox process and the Laplace mechanism. For each case, we specify the conditions to achieve $\ed$-DP under the $\alpha$-neighborhood. The proofs of the corresponding theorems and corollaries are provided in \cref{app:lemmas,app:kernel,app:lgcp,app:laplace}.

\subsection{Differential privacy for spatial point pattern}

Suppose we have an original spatial point pattern dataset $D=\{x_1,\ldots,x_n\}$ where each point $x_i$ lies in the compact spatial domain $S\subset \R^2$.
In the original context of DP \citep{dworkDifferentialPrivacy2006}, the neighboring datasets are defined as two datasets that differ by one record.
However, this definition may not be intuitive for spatial data because of the inherent spatially dependent structure within such data. The greater the distance between two points, the less their difference impacts the overall privacy. That is, the importance of privacy may depend on the local structure of the spatial data. Also, this discrepancy may lead to loss of utility if we use model-based synthesis for spatial point pattern data. To address these issues, we adopt the definition of DP under neighborhood by \citet{fangDifferentialPrivacyDneighbourhood2014}, which is more suitable for spatial data. 

\begin{definition}[$\alpha$-neighborhood]
\label{def:alpha-nb}
    Two point patterns $\allowbreak D=\{x_1,\ldots,x_{i-1},x_i,x_{i+1},\ldots,x_n\}$ and $\allowbreak D'=\{x_1,\ldots,x_{i-1},x_i',x_{i+1},\ldots,x_n\}$ are $\alpha$-neighboring, denoted as $D\nb D'$, when $d (x_i,x_i')\le \alpha$ for one pair of points $(x_i, x_i')$, where $d (\cdot,\cdot)$ is the distance function defined on $S$.
\end{definition}

\subsection{Poisson point synthesizers}
\label{subsec:PPS}

We define the Poisson point synthesizer as a randomized mechanism designed to generate synthetic point pattern datasets using a Poisson point process constructed from the original dataset $D$.

\begin{definition}
\label{def:pps}
A randomized mechanism $\M: N_\lf \to N_\lf$ is a \textit{point synthesizer} that takes original points $D \in N_\lf$ as input and outputs synthetic points $\M(D) \in N_\lf$. It is referred to as a Poisson point synthesizer (PPS) if  
\begin{equation}
    \M (D)\sim \Po (S, \lambda_D),
    \label{eq:pps}
\end{equation}
where $S$ denotes a spatial domain, and the intensity function $\lambda_D$ is derived from the original dataset $D \in N_\lf$.
\end{definition}

The differential privacy of point synthesizers under the $\alpha$-neighborhood is defined as follows.

\begin{definition}[DP for point synthesizers]
\label{def:dp_pps}
    The point synthesizer $\M$ satisfies $(\epsilon,\delta)$-differential privacy under the $\alpha$-neighborhood if, for all $D\nb D'\in N_\lf$ and $A \in \mathcal{N}_\lf$,
    \begin{equation}
      P (\M (D) \in A) \leq e^\epsilon P (\M (D') \in A) + \delta.
      \label{eq:dp_pps}
    \end{equation}
\end{definition}
Since the Poisson point process has a density with respect to the unit rate Poisson point process \citep{mollerStatisticalInferenceSimulation2003}, the condition \eqref{eq:dp_pps} for the Poisson point synthesizer can be rewritten as follows.

\begin{remark}[Density condition]
\label{remark:density_cond_dp}
    The PPS $\M$ satisfies $\ed$-differential privacy under the $\alpha$-neighborhood if, for all $D\nb D'\in N_\lf$, there exists $A_{D,D'}\in \mathcal{N}_\lf$ such that
    \begin{equation}
        e^{-\int_S \lambda_D (s)ds} \prod_{s\in \x}\lambda_D (s) \leq e^{\epsilon-\int_S \lambda_{D'}(s)ds} \prod_{s\in \x}\lambda_{D'}(s)
        \label{eq:dp_pps_density}
    \end{equation}
    holds for all $\x\in A_{D,D'}$ and $P (A_{D,D'})\ge 1-\delta$.
\end{remark}

One trivial example of the differentially private PPS is using a homogeneous Poisson point process as $\lambda_D(s) = |D|/|S|$ since relocating a single point does not change $\lambda_D(s)$.

\subsubsection{Kernel intensity estimation}
\label{sec:kernel}

We propose a Poisson point synthesizer utilizing the kernel intensity estimation, which satisfies $\alpha$-neighborhood differential privacy. The kernel intensity estimation is a common nonparametric approach to estimate the intensity function of a point process. We consider the edge-corrected kernel intensity estimator introduced by  \citet{diggleKernelMethodSmoothing1985} for the $\lambda_D$ of the PPS and refer to it as a kernel synthesizer. The kernel synthesizer, $\kernel(D)$, is defined as 
    \begin{gather}
        \kernel(D) \sim \Po (S, \lambda_D) \hbox{~ with~} \nonumber \\
        \lambda_D (s) = \sum_{x_i\in D}\frac{K_h (s-x_i)}{c_h (x_i)},
        \label{eq:edge_correction}
    \end{gather}
    where $K_h (\cdot)= \frac{1}{h^d} K (\cdot/h)$ is some kernel function with bandwidth $h$ on $S$ and $c_h (x_i)=\int_S K_h (u-x_i)du$ is the edge correction factor.

\begin{remark}
\label{remark:DP_bddkernel}
    When the kernel function has bounded support (e.g., Epanechnikov kernel), the $\kernel$ may not satisfy differential privacy. Let $S_D^0 = \{s \in S : \lambda_D(s) = 0\}$ be the subset of $S$ where the intensity function $\lambda_D$ is zero. If $S_D^0$ is not empty and there exists a synthetic point $s_0 \in S_D^0 \setminus S_{D'}^0$, the condition in the equation \eqref{eq:dp_pps_density} is violated, indicating that such a PPS does not satisfy differential privacy. Also note that kernel functions with a bounded support on an unbounded domain $S$ could give rise to such cases.
\end{remark}

To investigate a sufficient condition for DP of the kernel synthesizer, $\kernel(D)$, we use the Gaussian kernel function $K_h (s-x) = \frac{1}{h^d}\phi (\|s-x\|/h)$, where $\phi$ is the standard Gaussian density function.

\begin{theorem}
    \label{thm:kernel}
    The PPS $\kernel(D)$ with the intensity function given by \eqref{eq:edge_correction} satisfies $(\epsilon,\delta)$-DP under the $\alpha$-neighborhood if the following holds for some nonnegative integer $k$:
    \begin{gather*}
      P (Y \le k) \ge 1-\delta \hbox{~ and ~}\\
      \frac{2\alpha B+\alpha^2}{2h^2}+r_\alpha (h) \le \frac{\epsilon}{k},
    \end{gather*}
    where $Y$ is a Poisson random variable with mean $n$, $B=\max\{d (x,x'): x,x'\in S\}$, and $r_\alpha (h)=\max_{x,y\in S, d (x,y)\le \alpha}\left|\log\frac{c_h (x)}{c_h (y)}\right|$ is the maximum ratio of the edge correction factor for $\alpha$-neighboring points.
\end{theorem}

From the \cref{thm:kernel}, we can derive the required order of the perturbation size $\alpha$ with respect to the privacy parameters $\epsilon$ and $\delta$. For example, to achieve Scott's rule-of-thumb bandwidth of $O (n^{-1/5})$ \citep{scottMultivariateDensityEstimation2015} we need $O (n^{-7/5})$ order for $\alpha$. The details are provided in \cref{app:alpha_order}.

\subsection{Cox point synthesizers}
\label{subsec:CPS}

We propose another type of point synthesizer based on the Cox processes.

\begin{definition}
    \label{def:CPS}
    A randomized mechanism $\M: N_\lf \to N_\lf$ with a random intensity function $\lambda(\cdot)$ is a Cox point synthesizer if
    \begin{equation*}
        \M (D) \mid \lambda (\cdot) \sim \Po (S,\lambda (\cdot))
    \end{equation*}
    for the original dataset $D\in N_\lf$.
\end{definition}

Unlike Poisson point synthesizers, Cox point synthesizers introduce randomness into the intensity function $\lambda (\cdot)$. Here we assume that the intensity function implicitly depends on the dataset $D$ by the random vector $\THETA$. That is, $\M(D) \perp D \mid \THETA$ and $\M (D) \mid\THETA \sim \Po (S, \lambda_{\THETA})$.

\subsubsection{Log-Gaussian Cox process}

We now introduce a concrete example of a CPS based on the log-Gaussian Cox process \citep{mollerLogGaussianCox1998}. Specifically, we use the discretized (finite-dimensional) form of the latent Gaussian process $Y (s; \BETA)$, represented by the finite Gaussian random vector $\BETA$. For the discussion of CPS using LGCP, we further assume that the dataset $D$ follows a Cox process with the random intensity function $\lambda_{
\THETA}$, enabling us to handle the likelihood $p (D \mid \THETA)$

\begin{figure}
    \centering
    \begin{subfigure}{0.47\linewidth}
        \centering
        \includegraphics[width=\linewidth]{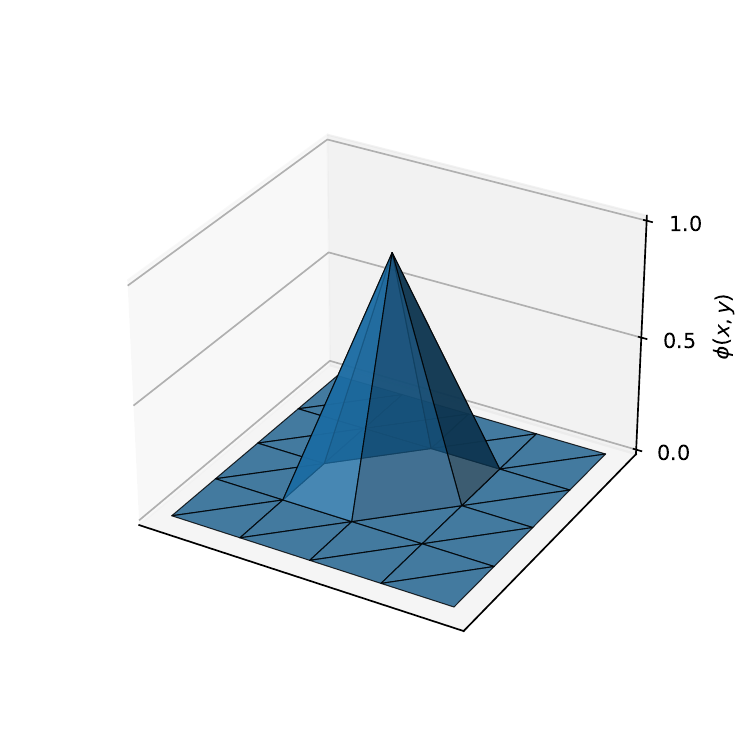}
        \caption{Triangular basis function.}
        \label{fig:triangularbasis}
    \end{subfigure}
    \begin{subfigure}{0.47\linewidth}
    \centering
        \includegraphics[width=\linewidth]{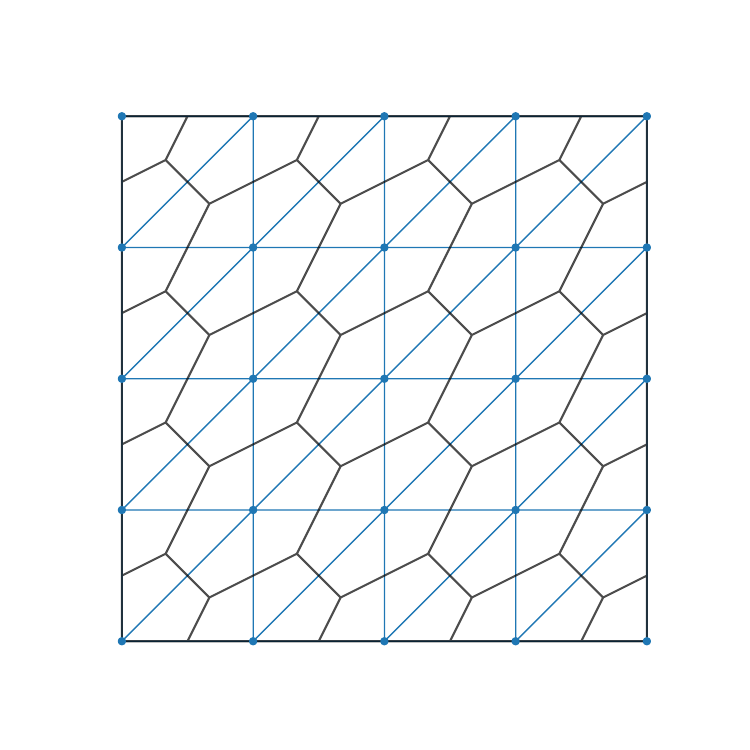}
        \caption{Triangulation (blue) and \\ its Voronoi dual mesh (black)}
        \label{fig:voronoi}
    \end{subfigure}
    \caption{Triangulation and basis function visualization.}
    \label{fig:triangulation_and_basis}
\end{figure}

Consider the regular tessellation of the spatial domain $S$ with $N$ grid points $t_1,\ldots,t_N$ such as the regular triangulation which divides each dimension of $S$ into $n_x$ and $n_y$ knot points (see \cref{fig:triangulation_and_basis}). If $S$ is rectangular, $S$ is divided into $2(n_x-1)(n_y-1)$ triangles. Then, we use the following intensity model:
\begin{equation}
    \log \lambda (s;\BETA) = Y (s;\BETA)= \lambda_0 + \sum_{i=1}^{N}\beta_i\phi_i (s),
    \label{eq:gcp}
\end{equation}
where $\lambda_0$ is the baseline intensity, $\BETA=(\beta_1,\ldots,\beta_N)$ is the random vector and $\{\phi_i (s)\}_{i=1}^N$ are basis functions. For $\phi_i (s)$, we consider a triangular piecewise linear basis function, which is a linear interpolation of the $i$th vertex of the triangle to the connected vertices (see \cref{fig:triangularbasis}) \citep{lindgrenExplicitLinkGaussian2011}. For the case of the linear network, the basis function is given by the linear interpolation of the two endpoints of the edge. We refer to this CPS as $\lgcp$.

For the distribution of the random vector $\BETA$, we consider the Bayesian approach. We assume that the prior distribution of $\BETA$ is the multivariate Gaussian distribution with zero mean and the covariance matrix $\Sigma$. We specify the following power covariance structure for $\Sigma$:
\begin{equation}
    \Sigma_{ij} = \sigma^2\exp\left (-\left (\frac{\|t_i-t_j\|}{l}\right)^p\right),
    \label{eq:covariance}
\end{equation}
where $\sigma^2$ is the variance, $l$ is the scale parameter, and the exponent $0<p\le 2$ determines the smoothness of the correlation function. 
We show that the exponent $p=2$ is required to achieve differential privacy for the two-dimensional spatial data in \cref{cor:equal_triangulation} when the spatial domain is square and the discretization of the domain is based on isosceles right triangles.

To implement $\lgcp$, we first sample $\BETA$ from the posterior distribution given the original dataset, $p (\BETA|D)$, using the following Bayesian formulation:
\begin{gather*}
    \sigma^2, l \sim \pi (\sigma^2, l)\\
    \BETA \sim \mathbf{N}(0, \Sigma (\sigma^2, l))
\end{gather*}
where $\pi(\sigma^2, l)$ represents the prior distribution over the parameters $\sigma^2$ and $l$. Using the sampled $\BETA$, we simulate the Poisson point process with $\lambda(s;\BETA)$ as the synthetic data.

To derive the DP condition for $\lgcp$, we introduce the following set of pairs of the triangles $T_i=(t_{i_1},t_{i_2},t_{i_3})$
\begin{equation}
    I_\alpha = \{(T_i, T_j) : d (T_i,T_j)\le \alpha, i < j\},
    \label{eq:pair}
\end{equation}
where $d (T_i, T_j)$ is the shortest distance between the two triangles. If $T_i$ and $T_j$ are adjacent, $d (T_i, T_j)$ is zero. The following theorem provides the DP condition for $\lgcp$.
\begin{theorem}
    \label{thm:gcp}
    Let $\lgcp$ be the Cox point synthesizer based on the log-Gaussian Cox process with intensity function \eqref{eq:gcp}. Assume that the dataset $D$ follows the same Cox process to construct $\lgcp$ and the domain $S$ is partitioned into right triangles. Then, $\lgcp$ is $(\epsilon, \delta)$-differentially private under the $\alpha$-neighborhood if
    \begin{equation}
        \delta \geq \frac{4}{\epsilon^2}\sum_{(i,j)\in I_\alpha}\sup_{\substack{\sum_{m=1}^{3} w_m = 1 \\ \sum_{m=1}^{3} v_m = 1}}\var\big (\eta_{i}(w) - \eta_{j}(v)\big),
        \label{eq:hat_basis}
    \end{equation}
    where $w =(w_1, w_2, w_3)$ and $v =(v_1, v_2, v_3)$ are weights for the triangle vertices, $I_\alpha$ is the set of neighboring triangle pairs defined in \eqref{eq:pair}, and $\eta_{i}(w) = \sum_{m=1}^{3} w_m \beta_{i_m}$ with $\beta_{i_m}$ as the basis function value at the $m$-th vertex of the $i$-th triangle.
\end{theorem}

Intuitively, $\eta_{i}(w)$ represents the weighted combination of the basis function values at the vertices of the $i$th triangle, with weights $w =(w_1, w_2, w_3)$ defining the interpolation over the triangle. We can set $\lambda_0 = |D|/|S|$ for simplicity since the choice of $\lambda_0$ does not affect the DP condition.

Next, we consider the specific case where the spatial domain is square $S=[0, B]^2$ and divided into $2N^2$ isosceles right triangles. By controlling the perturbation size $\alpha \le B/N\sqrt{2}$, we get the following corollary about the relationship between privacy and covariance function parameters. 

\begin{corollary}[Equal triangulation]
    \label{cor:equal_triangulation}
    Let $\lgcp$ be the Cox point synthesizer with the intensity function given by \eqref{eq:gcp} and \eqref{eq:covariance}. Under the equal triangulation of $S=[0,B]^2$ into $2N^2$ isosceles right triangles, $\lgcp$ is $\ed$-DP under the $\alpha$-neighborhood if the following holds:
    \begin{align}
        \alpha &\le \frac{B}{N\sqrt{2}} \nonumber \\
        \delta &\ge \frac{8B^p\sigma^2}{\epsilon^2 N^{p-2}l^p}(\sqrt{2}^p + 2^p + 6\cdot \sqrt{5}^p + 4\cdot \sqrt{8}^p) \nonumber \\
        &= \frac{544B^2\sigma^2}{\epsilon^2 l^2} \quad \text{for } p=2.
        \label{eq:corollary}
    \end{align}
\end{corollary}

\cref{cor:equal_triangulation} provides the relationship between the privacy parameters $\epsilon,\delta$ and the covariance function parameters for the log-Gaussian Cox process. Note $p=2$ in \eqref{eq:covariance} is appropriate to ensure that the parameter does not depend on the grid size $N$. If $p<2$, the privacy parameter $\delta$ increases as the grid size $N$ increases. This implies that the privacy level is not guaranteed for a finer grid. Therefore, we set $p=2$ in further analysis.

For the case of $p=2$, it is observed that the ratio of standard deviation to the scale parameter, $R = \sigma/l$ matters to determine the DP condition. During the posterior sampling, this ratio $R$ is fixed to the upper bound provided in \cref{cor:equal_triangulation}. We impose the prior on $l$ as $l \sim \pi_l$ so that $\sigma$ is determined by $\sigma = l\cdot R$.

The condition depends on the number of pairs of triangles, as well as the type of tessellation. If we use the square tessellation and square linear basis functions, the condition \eqref{eq:corollary} becomes $\alpha \le \sqrt{2}/N$ and $\delta \ge \frac{104B^2\sigma^2}{\epsilon^2 l^2}$ for $p=2$, which gives 5 times smaller requirement for the ratio $R$ compared to the equal triangulation.

The integral $\int_S \lambda (s) ds$ in the likelihood function is computationally intractable under the above formulation \eqref{eq:gcp}. To address this issue, we approximate the integral using the Voronoi dual mesh (see \cref{fig:voronoi}), as detailed in the \cref{app:voronoi} \citep{simpsonGoingGridComputationally2016}.

\subsubsection{Laplace mechanism}
\label{sec:lm}

The Laplace mechanism, introduced in the previous section, can be employed to the framework of CPS. For the Laplace mechanism CPS, we construct an intensity function using piecewise constant basis functions defined on the spatial domain $S$, which is partitioned into $N$ grid cells $\{S_i\}_{i=1}^N$, with each cell having an area of $|S_i|$. Specifically, the intensity function is defined as
\begin{equation}
    \label{eq:intensity_LM}
    \lambda (s; \GAMMA) = \sum_{i=1}^N \gamma_i \psi_i (s),
\end{equation}
where $\psi_i (s)$ is a piecewise constant function that equals $1/|S_i|$ if the point $s$ lies within the $i$th grid cell and $0$ otherwise. The parameter $\GAMMA=(\gamma_1,\ldots,\gamma_N)$ is given by 
\begin{equation}
    \label{eq:param_LM}
    \gamma_i = \max\left\{0, \sum_{x \in D}\psi_i (x) + {\rm Lap}\left (\frac{\Delta}{\epsilon}\right)\right\}
\end{equation}
where $\Delta=\max_{p \neq q} \left| \frac{1}{|S_p|} + \frac{1}{|S_q|} \right|$. Then, the following theorem demonstrates the DP of the Laplace mechanism CPS. The Laplace mechanism CPS ensures pure differential privacy. Indeed, it does not require the relaxation of DP using the $\alpha$-neighborhood.

\begin{theorem}
    \label{thm:laplace}
    Let $\lap$ be the Cox point synthesizer with the intensity function given by \eqref{eq:intensity_LM}. Then, $\lap$ is $(\epsilon,0)$-DP under the $\alpha$-neighborhood for all $0<\alpha \le B=\max\{d (x,x'): x,x'\in S\}$.
\end{theorem}

\subsection{Evaluation}

Synthetic data is evaluated using two metrics: (\romannumeral1) risk and (\romannumeral2) utility. The risk is quantified by the disclosure probability of individual points, which corresponds to the privacy budget $\epsilon$ in the differential privacy. The utility measures how much the synthetic data resembles the original data, which can be measured using various statistics and metrics. Specifically, we consider the K-function \citep{quickBayesianMarkedPoint2015} and the propensity mean squared error (pMSE) \citep{wooGlobalMeasuresData2009, snokeGeneralSpecificUtility2016} as the utility metrics.

The K-function is a common choice for analyzing spatial correlation of point patterns. We consider the nonparametric edge correction estimator for the inhomogeneous K-function \citep{baddeleySpatialPointPatterns2016}: 
\begin{equation}
    \hat{K}(r) = \frac{1}{|S|}\sum_{i=1}^{n}\sum_{j\ne i} \frac{\mathbf{1}(\|x_i-x_j\|\le r)}{\lambda (x_i)\lambda (x_j)p (x_i,d_{ij})}
\end{equation}
where $|S|$ is the area of the spatial domain $S$ and $\lambda (x)$ is the intensity function. The edge correction factor $p (x,d)$ is given by
\begin{equation*}
    p (x,d) = \frac{\ell (S \cap \partial b (x,d))}{2\pi d},
\end{equation*}
where $\partial b(x,d)$ is the circumference of a circle with radius $d$ centered at $x$ and $\ell (S \cap \partial b (x,d))$ is the length of $\partial b(x,d)$ intersecting $S$.
We compare the K-function of the original dataset with that of the synthetic datasets to evaluate their utility in capturing spatial correlation.

The pMSE is defined as the mean squared error of the propensity score, which represents the probability that a point is synthetic. To compute this, the original and synthetic datasets are concatenated with an indicator variable that identifies whether each point is synthetic. The propensity score is then estimated by classification models such as logistic regression or the tree-based models (e.g., CART, random forest) \citep{snokeGeneralSpecificUtility2016}.

However, the classification model is not appropriate for a point synthesizer in that only two variables-the coordinates of points-are available. Thus, instead of estimating the propensity score with the classification model, we use values of the intensity function to estimate the propensity score \citep{walderPrivacySpatialPoint2020}.
Let $D=(x_1,\ldots,x_n)$ and $\M (D)=(x_1',\ldots,x_m')$ be the observed and the synthetic datasets, respectively. Assume $P (x \in D) = \frac{n}{n+m}$ and $P (x \in \M (D)) = \frac{m}{n+m}$ naturally. Then, the propensity score is estimated by the normalized intensity function $\lambda_\text{norm}(s)$ as follows:
\begin{equation*}
    \hat{p}_i = P (x_i \in \M (D) | \lambda, \lambda') = \frac{\lambda'_\text{norm}(x_i)}{\lambda_\text{norm}(x_i) + \lambda'_\text{norm}(x_i)}
\end{equation*}
where $\lambda_\text{norm}(s) = \lambda (s) / \int_S \lambda (s)ds$ and $\lambda'_\text{norm}(s) = \lambda'(s) / \int_S \lambda'(s)ds$ are the normalized intensity functions of the observed and the synthetic datasets, respectively. The pMSE is then defined as
\begin{equation}
    \text{pMSE} = \frac{1}{n+m}\sum_{i=1}^{n+m}\left (\hat{p}_i - \frac{m}{n+m}\right)^2.
    \label{eq:pMSE}
\end{equation}

%% file: texs/experiments.tex
We conducted a simulation study to verify the risk and the utility of synthesized data for the point synthesizers. The study involves both simulated data generated from known intensity functions and real-world data on the linear network. For the CPS using the log-Gaussian Cox process, we used the Markov chain Monte Carlo posterior sampling using the No-U-Turn Sampler (NUTS) \citep{JMLR:v15:hoffman14a} to draw intensity samples from the posterior distribution. Also, we adopted the likelihood approximation using the Voronoi dual mesh. All experiments were implemented in Python with the \texttt{PyMC} library \citep{patilPyMCBayesianStochastic2010}.

We compared three methods: (1) the kernel estimation using a Gaussian kernel with edge correction (Kernel, $\kernel$), (2) the log-Gaussian Cox process method with triangulation (LGCP, $\lgcp$), and (3) the Laplace mechanism (Lap, $\lap$). For the Laplace mechanism, the granularity of the grid cells was set to match the grid size used in the log-Gaussian Cox process. Additionally, we examined the impact of different tessellation types on the utility of synthetic data, as detailed in \cref{sec:tessellation}. The results demonstrate that the choice of tessellation has minimal impact on the utility of the synthetic data.

\subsection{Simulated data}
We consider 4 types of true intensity functions $\lambda_i, i=1,2,3,4$ defined on two-dimensional squares $S_i\subset \R^2$ respectively. The intensity functions and spatial domains are given by
\begin{gather*}
    \lambda_1(x,y) = 10, ~~ \lambda_2(x,y) = e^{-\frac{x^2+y^2}{25}}, \\
    \lambda_3(x,y) = \frac{1}{2}+5e^{-(x-y)^2} \\
    \lambda_4(x,y) = 5e^{-\frac{(x-3)^2+(y-3)^2)}{2}} + 5e^{-\frac{(x+3)^2+(y+3)^2}{2}} \\
    S_1 = [0,1]^2, S_2 = [-10,10]^2, S_3 = [0,10]^2, S_4 = [-5,5]^2
\end{gather*}

\begin{figure}
    \centering
    \begin{subfigure}{0.45\columnwidth}
        \includegraphics[width=\linewidth]{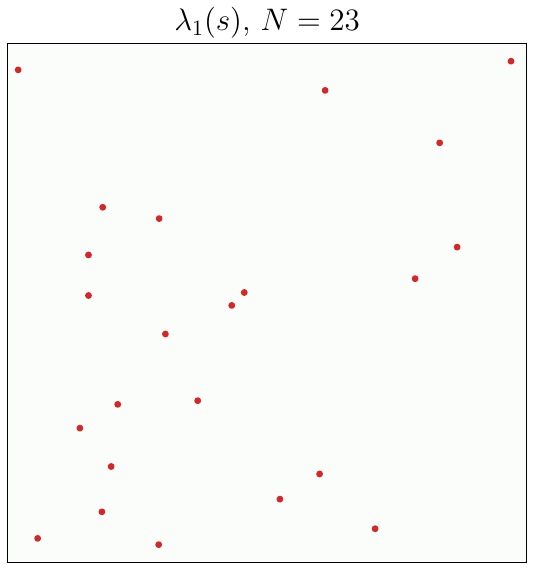}
    \end{subfigure}
    \begin{subfigure}{0.45\columnwidth}
        \includegraphics[width=\linewidth]{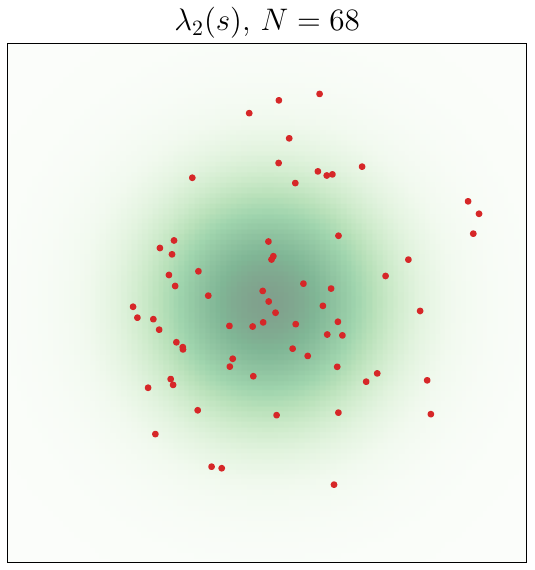}
    \end{subfigure}
    \begin{subfigure}{0.45\columnwidth}
        \includegraphics[width=\linewidth]{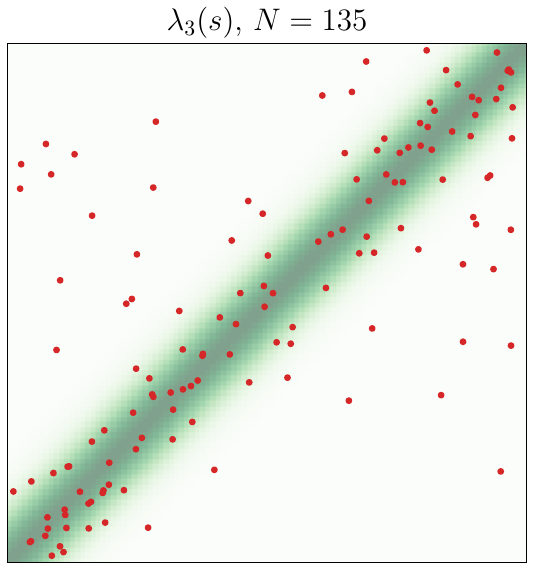}
    \end{subfigure}
    \begin{subfigure}{0.45\columnwidth}
        \includegraphics[width=\linewidth]{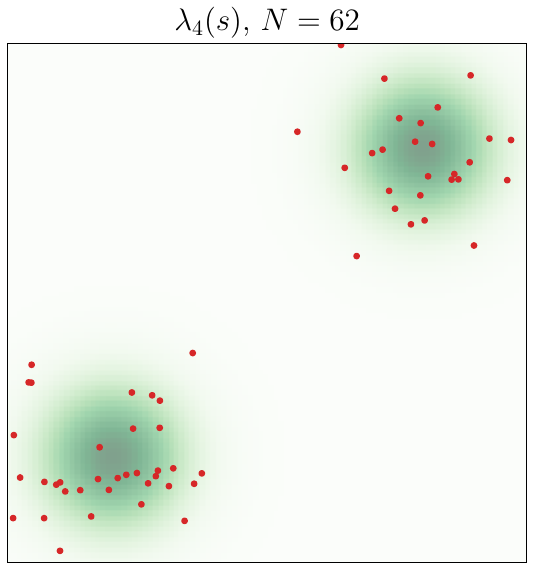}
    \end{subfigure}
    \caption{original dataset $D_{i,j}$ sampled from the intensity function $\lambda_i (x,y)$.}
    \label{fig:2d_intensity}
\end{figure}

The first case is a homogeneous Poisson process and the other cases are inhomogeneous Poisson processes with one cluster, one cluster along the diagonal, and two clusters, respectively. We sample $n_{\rm ori}=10$ datasets $D_{i,j}$ from each intensity function $\lambda_i(x,y)$ and generate 10 synthetic datasets $\M(D_{ij})_k$ for each dataset $D_{i,j}$. The privacy parameters are set to $\epsilon = 0.1, 1, 10$ and $\delta = 1/n$, where $n=|D_{i,j}|$ is the size of the original dataset. We use $n_x = n_y = 11$ knot points per dimension for grid granularity. The simulation results are summarized in \cref{tab:simulation}. The pMSE column presents the average pMSE and its standard deviation across the $n_{\rm syn}=10$ synthetic datasets. The npoints column shows the average number of synthetic points. The MISE column reports the averaged mean integrated squared relative error defined as
\begin{equation*}
    \resizebox{\hsize}{!}{$    \text{MISE} = \frac{1}{n_{\rm ori}n_{\rm syn}}\sum_{j=1}^{n_{\rm ori}}\sum_{k=1}^{n_{\rm syn}}\int_{\{r:\hat{K}^{\text{ori}}_{ij}(r) > 0\}} \left (\frac{\hat{K}^{\text{syn}}_{ijk}(r)}{\hat{K}^{\text{ori}}_{ij}(r)}-1\right)^2 dr.$}
\end{equation*}
Here, $\hat{K}^{\text{ori}}_{ij}(r)$ denotes the K-function estimate of $j$th sample original dataset from the intensity function $\lambda_i$ and $\hat{K}^{\text{syn}}_{ijk}(r)$ denotes the K-function estimate of the $k$th synthetic data generated from $D_{ij}$.

\begin{table}
\centering
\caption{Simulation results}
\label{tab:simulation}
\small
\begin{tabular}{lccccccc}
\toprule
$\lambda_i$ & $\epsilon$ & Method & pMSE & npoints & MISE \\ 
\midrule
\multirow{12}{*}{$\lambda_{1}$} & \multirow{4}{*}{0.1} & Ori &               - &    19.1 &               - \\
              &      & Kernel &  \textbf{0.003} &    20.2 &           0.306 \\
              &      & LGCP &           0.004 &    19.1 &           0.198 \\
              &      & Lap &           0.095 &   973.8 &  \textbf{0.025} \\
\cmidrule{2-6}
              & \multirow{4}{*}{1.0} & Ori &               - &    19.1 &               - \\
              &      & Kernel &  \textbf{0.002} &    20.0 &           0.197 \\
              &      & LGCP &           0.004 &    19.6 &           0.125 \\
              &      & Lap &           0.077 &   109.4 &  \textbf{0.022} \\
\cmidrule{2-6}
              & \multirow{4}{*}{10.0} & Ori &               - &    19.1 &               - \\
              &      & Kernel &  \textbf{0.003} &    20.8 &           0.280 \\
              &      & LGCP &           0.005 &    19.3 &           0.178 \\
              &      & Lap &            0.05 &    27.5 &  \textbf{0.033} \\
\cmidrule{1-6}
\cmidrule{2-6}
\multirow{12}{*}{$\lambda_{2}$} & \multirow{4}{*}{0.1} & Ori &               - &    77.6 &               - \\
              &      & Kernel &           0.076 &    78.5 &           0.850 \\
              &      & LGCP &  \textbf{0.076} &    77.2 &           0.865 \\
              &      & Lap &           0.519 &  1014.4 &  \textbf{0.364} \\
\cmidrule{2-6}
              & \multirow{4}{*}{1.0} & Ori &               - &    77.6 &               - \\
              &      & Kernel &   \textbf{0.07} &    78.3 &           0.836 \\
              &      & LGCP &           0.076 &    77.2 &           0.882 \\
              &      & Lap &           0.273 &   154.7 &  \textbf{0.264} \\
\cmidrule{2-6}
              & \multirow{4}{*}{10.0} & Ori &               - &    77.6 &               - \\
              &      & Kernel &  \textbf{0.037} &    76.9 &  \textbf{0.795} \\
              &      & LGCP &           0.076 &    76.9 &           0.929 \\
              &      & Lap &           0.212 &    84.3 &           1.148 \\
\cmidrule{1-6}
\cmidrule{2-6}
\multirow{12}{*}{$\lambda_{3}$} & \multirow{4}{*}{0.1} & Ori &               - &   128.5 &               - \\
              &      & Kernel &           0.052 &   129.2 &  \textbf{0.102} \\
              &      & LGCP &  \textbf{0.052} &   129.3 &           0.114 \\
              &      & Lap &           0.106 &  1076.6 &           0.223 \\
\cmidrule{2-6}
              & \multirow{4}{*}{1.0} & Ori &               - &   128.5 &               - \\
              &      & Kernel &           0.052 &   129.1 &  \textbf{0.100} \\
              &      & LGCP &  \textbf{0.052} &   128.9 &           0.108 \\
              &      & Lap &           0.062 &   195.3 &           0.135 \\
\cmidrule{2-6}
              & \multirow{4}{*}{10.0} & Ori &               - &   128.5 &               - \\
              &      & Kernel &           0.049 &   127.7 &  \textbf{0.097} \\
              &      & LGCP &           0.052 &   129.9 &           0.110 \\
              &      & Lap &   \textbf{0.03} &   132.8 &           0.553 \\
\cmidrule{1-6}
\cmidrule{2-6}
\multirow{12}{*}{$\lambda_{4}$} & \multirow{4}{*}{0.1} & Ori &               - &    57.3 &               - \\
              &      & Kernel &  \textbf{0.129} &    58.2 &           5.087 \\
              &      & LGCP &            0.13 &    56.7 &           4.258 \\
              &      & Lap &           0.497 &  1043.7 &  \textbf{0.174} \\
\cmidrule{2-6}
              & \multirow{4}{*}{1.0} & Ori &               - &    57.3 &               - \\
              &      & Kernel &            0.13 &    58.4 &           4.880 \\
              &      & LGCP &   \textbf{0.13} &    58.1 &           5.184 \\
              &      & Lap &           0.282 &   144.0 &  \textbf{0.347} \\
\cmidrule{2-6}
              & \multirow{4}{*}{10.0} & Ori &               - &    57.3 &               - \\
              &      & Kernel &  \textbf{0.108} &    57.7 &           5.781 \\
              &      & LGCP &           0.128 &    57.6 &           4.851 \\
              &      & Lap &           0.127 &    64.8 &  \textbf{0.673} \\
\bottomrule
\end{tabular}
\end{table}

The results indicate that the $\kernel$ and $\lgcp$ methods outperform $\lap$ in terms of pMSE. However, $\lap$ demonstrates superior utility regarding MISE when the original datasets are from inhomogeneous Poisson processes and the points exhibit clustering tendencies. When comparing the number of synthetic points to the original dataset (Ori), $\lap$ consistently generates a higher quantity than the other methods, particularly when the privacy budget is small. In addition, both the $\kernel$ and $\lgcp$ methods exhibit robustness with respect to the privacy budget $\epsilon$, as their pMSE and MISE remain largely unaffected by variations in $\epsilon$.

\subsection{Extension to linear networks}

To further evaluate our methods on the linear network, we applied them to real-world data from the city of Chicago\footnote{Retrieved from Chicago Data Portal. The data pertains to crimes reported in the year 2023 and is spatially restricted to the bounding box: $[-87.66, -87.655] \times [41.80, 41.805]$ using the WGS84 coordinate reference system.}, consisting of the locations of the reported crime incidents. To ensure the integrity of our analysis, we preprocessed the data by removing duplicates, guaranteeing the uniqueness of each point in the dataset. We compared the two CPS $\lgcp$ and $\lap$ with the original dataset. 

For the $\lgcp$ method, we employed the exponential correlation function with $\alpha=1$ to utilize the resistance metric (detailed in \cref{app:linear_network}). To prepare the linear network for two CPSs, we discretized the network by dividing each line segment into smaller subsegments. We ensured that each subsegment's length did not exceed the specified resolution of $r = 50$ meters. The privacy parameters were set to $\epsilon = 0.1, 1.0, 10.0$ and $\delta = 1/n$. Given that the true intensity function of the real-world data was unknown, we focused our utility evaluation on the MISEs of the K-function. Specifically, we employed the homogeneous K-function for this assessment. The details of the K-function application on the linear network are elaborated in \cref{sec:kfunctionlinnet}.

\begin{figure}[ht!]
    \centering
    \begin{subfigure}{0.45\columnwidth}
        \includegraphics[width=\columnwidth]{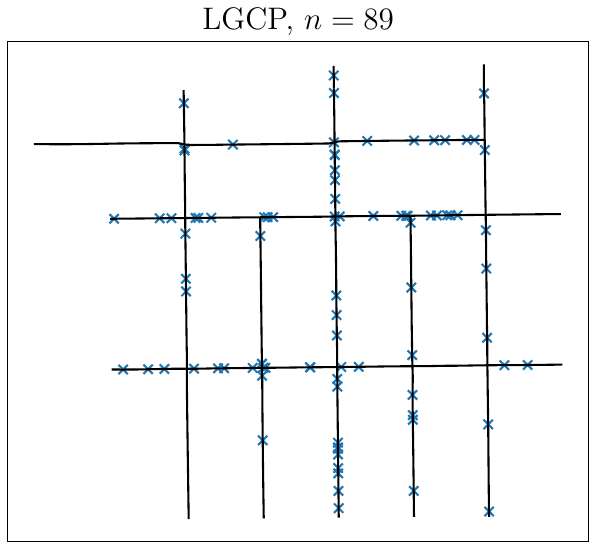}
    \end{subfigure}
    \begin{subfigure}{0.45\columnwidth}
        \includegraphics[width=\columnwidth]{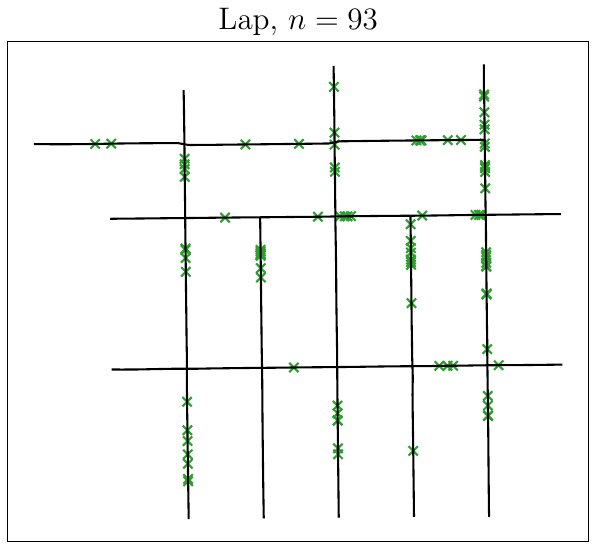}
    \end{subfigure}
    \begin{subfigure}{0.45\columnwidth}
        \includegraphics[width=\columnwidth]{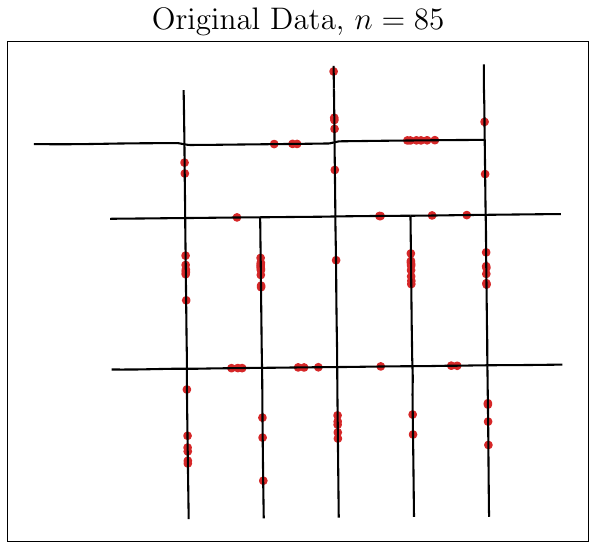}
    \end{subfigure}
    \caption{Synthetic Chicago crime data with $\epsilon=1.0$. \\
    Upper left: $\lgcp$, Upper right: $\lap$, Lower: Original dataset}
    \label{fig:linnet_result}
\end{figure}

An example of synthetic data is illustrated in \cref{fig:linnet_result}, with a summary of the results presented in \cref{tab:simulation_linnet}. The Laplace mechanism $\lap$ consistently outperforms $\lgcp$ in terms of MISE across all privacy budgets. This is evident at lower privacy budgets, indicating that $\lap$ produces more utilizable synthetic data under stricter privacy constraints. This is because $\lgcp$ failed to capture some clustered points (e.g. lower left side of the network in \cref{fig:linnet_result}) compared to $\lap$.

However, the improved utility of $\lap$ comes with a notable trade-off: it generates a significantly higher number of synthetic points than $\lgcp$. For instance, at $\epsilon = 0.1$, $\lap$ produces an average of 1443.6 points, while $\lgcp$ generates only 86.4 points, which is much closer to the original dataset size of 85 points. As demonstrated in our simulation study, $\lap$ exhibits superior utility since the original dataset displays clustering tendencies.

\begin{table}[ht!]
\centering
\caption{Simulation results for Chicago crime data. Original number of points is $|D|=85$. 30 synthetic datasets were generated for each $\epsilon$.}
\label{tab:simulation_linnet}
\small
\begin{tabular}{ccccc}
\toprule
$\epsilon$ & Method & MISE & npoints \\ 
\midrule
\multirow{2}{*}{0.1} & LGCP & 24.267 & 86.4 (10.57) \\
 & Lap & \textbf{8.288} & 1443.6 (182.15) \\
\cmidrule{1-4}
\multirow{2}{*}{1.0} & LGCP & 18.827 & 87.7 (12.42) \\
 & Lap & \textbf{4.013} & 209.9 (24.55) \\
\cmidrule{1-4}
\multirow{2}{*}{10.0} & LGCP & 8.284 & 83.8 (11.29) \\
 & Lap & \textbf{3.768} & 91.7 (9.19) \\
\bottomrule
\end{tabular}
\end{table}

%% file: texs/conclusion.tex
In this paper, we proposed a method to synthesize spatial point pattern data within the differential privacy framework. We introduced the Poisson point synthesizer and the Cox point synthesizer which generate synthetic points by simulating the Poisson point processes and Cox processes, respectively.
We provided three example classes of synthesizers based on how to handle an intensity function: kernel intensity estimator, the log-Gaussian Cox process (LGCP) and the Laplace mechanism. We provided some theoretical background that each synthesizer class satisfies the differential privacy under some conditions. The kernel method achieves DP by bounding the bandwidth. For the LGCP method, DP is achieved by bounding the covariance parameter ratio $R=\sigma/l$ and the Laplace mechanism naturally achieves pure DP.
Through a simulation study, we verified the utility and privacy guarantees of the proposed method. The kernel and LGCP methods showed robustness to the privacy budget $\epsilon$, as their propensity score mean squared error (pMSE) and mean integrated squared relative error of the K-function remained relatively stable despite changes in $\epsilon$. In contrast, the Laplace mechanism generates a substantially higher number of synthetic points than the proposed method, especially when the privacy budget is small. When applied to linear networks, the Laplace mechanism consistently demonstrates its superior utility to the LGCP method in terms of MISE. However, this improved utility comes with a trade-off, as the Laplace mechanism tends to generate a substantially higher number of synthetic points. These findings underscore the importance of balancing both utility and data volume when selecting a synthetic data generation method for privacy-preserving spatial point pattern analysis.

There are several directions for future research. Improving the stability of the Poisson point synthesizer remains a question, particularly for the Laplace mechanism which tends to generate an excessive number of points. While thinning might address this issue, determining the appropriate thinning probability requires further investigation. The proposed method could be extended to spatio-temporal domains. The log-Gaussian Cox process method could be adapted by incorporating suitable spatio-temporal covariance functions, while the kernel method may be extended with additional computations to verify differential privacy conditions in this context. Expanding the Poisson point synthesizer to handle marked point patterns, where each point is associated with additional variables (e.g. crime type), could broaden the applicability of this approach. Integrating methods such as the Bayesian marked point process model proposed by \citet{quickBayesianMarkedPoint2015} into the DP framework could broaden the applicability of our approach as well. These advancements would further establish the utility of differentially private synthesis methods for spatial point patterns across a wide range of real-world applications.

%% file: texs/appendicies.tex
\section{Lemmas}
\label{app:lemmas}

We first introduce lemmas with proofs which will be used for the proofs of the theorems in \cref{subsec:CPS}.

\begin{lemma}
    \label{lemma:dp_gcp}
    Suppose a Cox point synthesizer $\M(D)$ has an intensity function implicitly dependent on the dataset $D$. That is, the intensity function $\lambda(s;\THETA)$ is parametrized with some random vector $\THETA=(\theta_1,\ldots, \theta_N)$ so that $\M(D) \perp D \mid \THETA$. Then, $\M(D)$ is $\epsilon$-differentially private point synthesizer under the $\alpha$-neighborhood if the following condition holds: 
    \begin{align}
        p(\THETA | D) \le e^\epsilon p(\THETA | D') \quad \forall D\nb D'
        \label{eq:dp_gcp}
    \end{align}
\end{lemma}

\begin{proof}
    For the point process $\X=\M(D)$, we have for $A\in \mathcal{N}_\lf$
  \begin{align*}
    P(\X\in A|D) &= \int P(\X\in A|\THETA,D)p(\THETA|D)d\THETA\\
    &= \int P(\X\in A|\THETA)p(\THETA|D)d\THETA\\
    &\le \int P(\X\in A|\THETA) e^\epsilon p(\THETA|D')d\THETA=e^\epsilon P(\X\in A|D').
  \end{align*}
  The second equality holds since $\X$ is independent of $D$ given the intensity function $\lambda(s;\THETA)$. 

\end{proof}

\begin{lemma}
    \label{lemma:dp_gcp_posterior}
    Assume that the dataset $D$ follows the log-Gaussian Cox process with the intensity function $\lambda_{\BETA}(s)$.
    Then,
    \begin{equation}
        p(\BETA|D) \le e^\epsilon p(\BETA|D')\quad \forall D\nb D'
    \end{equation}
    holds if equation \eqref{lemma:eq:dp_gcp2} holds for all $x,x'\in S$ with $d(x,x')\le \alpha$,
    \begin{equation}
        |\log\lambda_{\BETA}(x)-\log\lambda_{\BETA}(x')| \le \frac{\epsilon}{2}.
        \label{lemma:eq:dp_gcp2}
    \end{equation}
\end{lemma}

\begin{proof}
    Let $D=(x_1,\ldots,x_n)$ and $D'=(x_1,\ldots,x_{n-1}, x_n')$ be the $\alpha$-neighboring datasets for the last element without loss of generality. From the assumption, the likelihood for the dataset $D$ is given by
    \begin{equation*}
        p(D|\BETA) = \exp\left(-\int_S \lambda_{\BETA}(s)ds\right) \prod_{i=1}^n \lambda_{\BETA}(x_i).
    \end{equation*}
Then, we have
    \begin{align*}
        \frac{p(\BETA|D)}{p(\BETA|D')} &= \frac{p(D| \BETA)p(\BETA)/P(D)}{p(D'| \BETA)p(\BETA)/P(D')} = \frac{p(D|\BETA)\int P(D'|\BETA)p(\BETA)d\BETA}{p(D'|\BETA)\int P(D|\BETA)p(\BETA)d\BETA} \\
        &=\frac{\lambda_{\BETA}(x_n)}{\lambda_{\BETA}(x_n')} \times \frac{\int \exp\left(-\int_S \lambda_{\BETA}(s)ds\right) \prod_{i=1}^n \lambda_{\BETA}(x_i) \frac{\lambda_{\BETA}(x_n')}{\lambda_{\BETA}(x_n)} p(\BETA) d\BETA}{\int \exp\left(-\int_S \lambda_{\BETA}(s)ds\right) \prod_{i=1}^n \lambda_{\BETA}(x_i)p(\BETA) d\BETA}\\
        &= \exp(\log \lambda_{\BETA}(x_n)-\log \lambda_{\BETA}(x_n')) \times \frac{\int P(D|\BETA)p(\BETA) \exp(\log \lambda_{\BETA}(x_n')-\log \lambda_{\BETA}(x_n))  d\BETA}{\int P(D|\BETA)p(\BETA) d\BETA}.
    \end{align*}
    From 
    \begin{equation*}
        |\log \lambda_{\BETA}(x_n) -  \log \lambda_{\BETA}(x_n')| \le \epsilon/2,
    \end{equation*}
    we have
    \begin{equation*}
        \frac{p(\BETA|D)}{p(\BETA|D')} \le e^{\epsilon/2} \times
        e^{\epsilon/2} \frac{p(D)}{p(D)} = e^\epsilon.
    \end{equation*}
    Thus, we get the desired result.
\end{proof}


\section{Proof of \cref{thm:kernel}}
\label{app:kernel}

\begin{proof}
    Let $D=\{x_1,\ldots,x_n\}$ and $D'=\{x_1,\ldots,x_{n-1},x_n'\}$ be $\alpha$-neighboring datasets. First, we consider the following $\epsilon$-DP condition for the synthetic dataset $\x=\{s_1,\ldots, s_k\}$:
    \begin{equation}
        \frac{p_D(\x)}{p_{D'}(\x)} = \exp\left(\int_S \lambda_{D'}(s)-\lambda_D(s) ds\right) \prod_{s\in \x}\frac{\lambda_D(s)}{\lambda_{D'}(s)} \le e^\epsilon,
        \label{eq:dp_pps_density_eps}
    \end{equation}
    where $p_D$ is the density of $\M_{\rm kern}(D)$ with respect to the unit rate Poisson point process. Note the first term in \eqref{eq:dp_pps_density_eps} becomes 1 since 
    \begin{equation*}
        \int_S \lambda_D(s)-\lambda_{D'}(s)ds = \int_S \frac{K_h(s-x_n)}{c_h(x_n)}ds - \int_S \frac{K_h(s-x_n')}{c_h(x_n')}ds = 0.
    \end{equation*}
    Thus, it suffices to bound $\prod_{s\in \x}\lambda_D(s)/\lambda_{D'}(s)$ for all $\x\in N_\lf$ by $e^{\epsilon}$, which leads to a following sufficient condition
  \begin{align*}
    &\prod_{i=1}^k \frac{\lambda_D(s_i)}{\lambda_{D'}(s_i)} \le e^\epsilon \\
    &\Leftarrow \frac{\lambda_D(s)}{\lambda_{D'}(s)} \le e^{\epsilon/k}\quad \forall s\in S\\
    &\Leftarrow \frac{K_h(s-x_n)}{K_h(s-x_n')} \le e^{\epsilon/k-r_\alpha(h)}\quad \forall s\in S,
  \end{align*}
  where $r_\alpha (h)=\max_{x,y\in S, d (x,y)\le \alpha}\left|\log\frac{c_h (x)}{c_h (y)}\right|$.

    For the Gaussian kernel $K=\phi$,
    \begin{align*}
      \frac{K_h(s-x)}{K_h(s-x')} &= \exp\left(-\frac{1}{2h^2}\left(\|s-x\|^2-\|s-x'\|^2\right)\right) \\
      &= \exp\left(\frac{1}{2h^2}(2(s-x)+(x-x'))^\top(x-x')\right) \\
      &\le \exp\left(\frac{2\alpha\|s-x\|+\alpha^2}{2h^2}\right)\\
      &\le \exp\left(\frac{2\alpha B+\alpha^2}{2h^2}\right)
    \end{align*}
    holds for $B=\max\{d(x,x'): x,x'\in S\}$. Thus, the condition \eqref{eq:dp_pps_density_eps} is satisfied for $\x \in N_\lf$ if
    \begin{equation*}
        \frac{2\alpha B+\alpha^2}{2h^2}+r_\alpha(h) \le \frac{\epsilon}{m}, \quad \forall m\in \{0\}\cup\N.
    \end{equation*}

    Let $N_\lf^k=\{\x\in N_\lf: n(\x)\le k\}$ be the set of locally finite point patterns with at most $k$ points. Then for some nonnegative integer $k$,
    \begin{equation*}
        \frac{2\alpha B+\alpha^2}{2h^2}+r_\alpha(h) \le \frac{\epsilon}{k} \Rightarrow \frac{p_D(\x)}{p_{D'}(\x)} \le e^{\epsilon}, \quad \forall \x\in N_\lf^k,
    \end{equation*}
    where $p_D(\x)$ is the density of the PPS $\M(D)$ at $\x$ with respect to the unit rate Poisson point process. Since 
    \begin{equation*}
        P(\x\in N_\lf^k) = P(Y\le k) \ge 1-\delta
    \end{equation*}
    for $Y\sim \mathrm{Poisson}(n)$, we get the following relationship:
    \begin{align*}
        P(\M(D) \in F) &= P(\M(D) \in F \cap N_\lf^k) + P(\M(D) \in F \cap {N_\lf^k}^c) \\
        &\le P(\M(D) \in F \cap N_\lf^k) + \delta \\
        &\le e^\epsilon P(\M(D') \in F \cap N_\lf^k) + \delta \\
        &\le e^\epsilon P(\M(D') \in F) + \delta,
    \end{align*}
    which is the desired result.
\end{proof}

\subsection{Derivation of the perturbation order}
\label{app:alpha_order}

First, we consider the relationship between $k$ and $\delta$. For some $s>0$, let $\delta$ be given by the order of $O(1/n^s)$ such that 
\begin{align*}
    P(Y\le k) &\ge 1-\delta, \\
    O(1/n^s) = \delta &\ge 1- P(Y\le k) = P(Y>k) = P(Y\ge k-1).
\end{align*}
From the Chernoff bound, we get
\begin{align*}
    P(Y\ge k-1) &\le \inf_{t>0} M_Y(t)e^{-t(k-1)} \\
    &= \inf_{t>0} \exp\left(n(e^t-1)-t(k-1)\right) \\
    &= \exp\left(n(\frac{k-1}{n}-1)-\log(\frac{k-1}{n})(k-1)\right) \\
    &= e^{-n}\left(\frac{en}{k-1}\right)^{k-1} \le O(1/n^s),
\end{align*}
where $M_Y(t)$ is the moment generating function of $Y\sim {\rm Poisson}(n)$ and the infimum is minimized at $t=\log((k-1)/n)$.

Next, we can derive the order of $k$ as $O(n+\sqrt{sn\log n})$ as follows:
\begin{align*}
    O(-s\log n) &\ge -n + (k-1)\log\left(\frac{en}{k-1}\right) \\
    &\ge -n + (n+t) \log\left(\frac{en}{n+t}\right) \quad (\text{by } k-1=n+t)\\
    &\ge -n + (n+t)\left(1-\frac{t}{n+t} - \frac{1}{2}\cdot\left(\frac{t}{n+t}\right)^2 + O(n^{-3})\right), \\
    \frac{t^2}{2n} &\ge O(s\log n).
\end{align*}
Here setting $k-1=n+t$ for $t>0$ is justified by the small privacy parameter $\delta$. Since $r_\alpha(h) $ approaches zero for sufficiently small $\alpha$, the theorem implies that $h \ge O(\sqrt{\frac{k\alpha}{\epsilon}})$ is required for the given $\epsilon$ and $\alpha$. To achieve Scott's rule-of-thumb bandwidth of $ O(n^{-1/5}) $ \citep{scottMultivariateDensityEstimation2015}, the order of $\alpha$ needs to be $O(n^{-7/5})$.

\section{Proof of \cref{thm:gcp}}
\label{app:lgcp}

\begin{proof}
    \cref{lemma:dp_gcp_posterior} gives the sufficient condition for the $\epsilon$-DP of the synthesizer $\M(D)$ as
    \begin{equation*}
        |\log \lambda_{\BETA}(x) - \log \lambda_{\BETA}(x')| \le \frac{\epsilon}{2}
    \end{equation*}
    for all $x,x'\in S$ with $d(x,x')\le \alpha$. From the following relationship,
    \begin{align*}
        P\left(\max_{(i,j)\in I_\alpha}|\eta_i(w)-\eta_j(v)|\le \epsilon/2\right) 
        &\le P\left(|\log\lambda_{\BETA}(x)-\log\lambda_{\BETA}(x')|\le \epsilon/2\right) \\
        &\le P\left(\log\frac{p(\M(D))}{p(\M(D'))} \le \epsilon\right),
    \end{align*}
    where $p(\M(D))$ is the probability density of the synthetic points $\M(D)$, it suffices to show that
    \begin{equation*}
        P\left(\max_{(i,j)\in I_\alpha}|\eta_{i}(w)-\eta_{j}(v)| \le \frac{\epsilon}{2}\right) \ge 1-\delta
    \end{equation*}
    for the $\ed$-DP of $\M(D)$ under the $\alpha$-neighborhood. 
    
    Suppose $x$ and $x'$ are located in the triangles $T_i$ and $T_j$, respectively. Since the value of the basis function at each point is given by the linear interpolation of the vertices of the triangle, we can express the difference of log intensities between $x$ and $x'$ as
    \begin{align*}
        \log \lambda_{\BETA}(x) - \log \lambda_{\BETA}(x') &= \lambda_0 + \sum_{i=1}^{N}\beta_i\phi_i(x) - \lambda_0 - \sum_{i=1}^{N}\beta_i\phi_i(x') \\
        &= \sum_{k=1}^{3}\beta_{i_k}\phi_{i_k}(x) - \sum_{k=1}^{3}\beta_{j_k}\phi_{j_k}(x') \\
        &= \sum_{k=1}^{3}w_k\beta_{i_k} - \sum_{k=1}^{3}v_k\beta_{j_k},
    \end{align*}
    where $w=(w_1,w_2,w_3)$ and $v=(v_1,v_2,v_3)$ are the basis function values at the vertices of the triangles $T_i$ and $T_j$ respectively. Here, $\sum_{k=1}^{3}w_k = \sum_{k=1}^{3}v_k = 1$.

    Let $\Gamma_{w,v} = (\eta_{i}(w)-\eta_{j}(v))_{(i,j)\in I_\alpha}$, where $\eta_{i}(w) = \sum_{m=1}^{3}w_m\beta_{i_m}$. Then, for any $w$ and $v$, $\Gamma_{w,v}$ is a Gaussian random vector. Denote the covariance matrix by $\Sigma_{w,v}$. Putting $\Gamma_{w,v} = \Sigma_{w,v}^{1/2}Z$, where $Z\sim\mathcal{N}(0,I)$ and $\Sigma_{w,v}=Q\Lambda Q^\top$, we have
    \begin{align*}
        P(\sup_{w,v} \max_{(i,j)\in I_\alpha}|\eta_{i}(w)-\eta_{j}(v)| \le \frac{\epsilon}{2}) &= P(\sup_{w,v} \|\Gamma_{w,v}\|_\infty \le \frac{\epsilon}{2}) \\
        &\ge P(\sup_{w,v} \|\Gamma_{w,v}\|_2 \le \frac{\epsilon}{2}) \\
        &= P(\sup_{w,v} \|Q\Lambda^{1/2}Q^\top Z\|_2^2 \le \frac{\epsilon^2}{4}) \\
        &= P(\sup_{w,v} \|\Lambda^{1/2}W\|_2^2 \le \frac{\epsilon^2}{4}) \quad (\text{Putting } W=Q^\top Z)\\
        &= P(\sup_{w,v} \sum_{(i,j)\in I_\alpha}\lambda_{ij}W_{ij}^2 \le \frac{\epsilon^2}{4}),
    \end{align*}
    where the second inequality holds from the relationship between the $l_2$ and $l_\infty$ norms. Here $\lambda_{ij}$ is the eigenvalue of the covariance matrix $\Sigma_{w,v}$, and $W_{ij}$ is the standard normal random variable. 

    Since $\sum_{(i,j)\in I_\alpha}\lambda_{ij}W_{ij}^2$ is the sum of the independent chi-square random variables, from the Markov inequality, we have
    \begin{equation*}
        P(\sup_{w,v} \max_{(i,j)\in I_\alpha}|\eta_{i}(w)-\eta_{j}(v)| \le \frac{\epsilon}{2}) \ge 1 - \frac{4}{\epsilon^2}\sum_{(i,j)\in I_\alpha}\sup_{w,v}\var(\eta_{i}(w)-\eta_{j}(v)).
    \end{equation*}



\end{proof}

\subsection{Proof of \cref{cor:equal_triangulation}}

\begin{proof}
    First, observe that the term $\sup_{w,v} \var(\eta_{i}(w) - \eta_{j}(v))$ reaches its maximum when $w = e_i$ and $v = e_j$, where $e_i \in \mathbb{R}^3$ is the unit vector, and $i$ and $j$ are the indices of the pair of vertices in the triangles that are farthest apart. By setting $\alpha \le \frac{B}{N\sqrt{2}}$, we can define $I_\alpha$ as the set of pairs of triangles that are adjacent by either a vertex or a side.

    Then, we get $N^2$ pairs for $d=\sqrt{2}B/N$, $(N-1)^2$ pairs for $d=2B/N$, $6N(N-1)$ pairs for $d=\sqrt{5}B/N$, and $4(N-1)^2$ pairs for $d=2\sqrt{2}B/N$, where $d$ is the maximum distance between the vertices of the triangles. Combining the results, we have
    \begin{align*}
        \sum_{(i,j)\in I_\alpha}\sup_{w,v}\var(\eta_{i}(w)-\eta_{j}(v)) 
        &= N^2 \cdot 2\sigma^2 (1-\exp(-(\sqrt{2}B/Nl))^p) \\
        &+ (N-1)^2 \cdot 2\sigma^2 (1-\exp(-(2B/Nl))^p) \\
        &+ 6N(N-1) \cdot 2\sigma^2 (1-\exp(-(\sqrt{5}B/Nl))^p) \\ &+ 4(N-1)^2 \cdot 2\sigma^2 (1-\exp(-(2\sqrt{2}B/Nl))^p) \\
        &\le \frac{2N^2B^p\sigma^2}{N^p l^p}(\sqrt{2}^p + 2^p + 6\cdot \sqrt{5}^p + 4\cdot \sqrt{8}^p)
    \end{align*}
    using $1-\exp(-x) \le x$. Putting this into the DP condition at \cref{thm:gcp}, we get the desired result. The results for the square tessellation can be proven similarly.
\end{proof}

\subsection{Likelihood approximation using the Voronoi dual mesh}
\label{app:voronoi}

For each vertex of the triangles, we construct a Voronoi dual cell by connecting the midpoints of its connected edges. Given the following intensity structure,
\begin{equation}
    \lambda (s) = \exp (\lambda_0 + \sum_{i=1}^N \beta_i \phi_i (s))
\end{equation}
we approximate the integral as:
\begin{equation}
    \int_S \lambda (s) ds \approx \sum_{i=1}^N |\alpha_i|  \exp (\lambda_0 + \sum_{j=1}^N \beta_j \phi_j (x_i))
\end{equation}
where $\alpha_i$ denotes the area of the $i$-th dual cell. Define $P_{ij} = \phi_j (x_i)$ as the projection matrix for the original dataset $D=(x_1,\ldots,x_n)$ onto the basis functions, and set

\begin{align*}
    \log\boldsymbol{\eta} &=(\w^\top, \w^\top P^\top)^\top,\\
    \boldsymbol{\alpha} &=(\alpha_1,\ldots,\alpha_N,\mathbf{0}_n^\top)^\top, \\ 
    \y &=(\mathbf{0}_N^\top, \mathbf{1}_n^\top)^\top
\end{align*} 

where $\w = \BETA + \mathbf{1}_N\lambda_0$. With these definitions, the approximate log-likelihood can be written as
\begin{equation*}
    \log p (\{x_1,\ldots,x_n\}|\lambda) = \sum_{i=1}^{N+n} y_i \log \eta_i - \alpha_i\eta_i.
\end{equation*}
Here $N$ is the number of vertices of the triangles and $n$ is the number of the original points.

\section{Proof of \cref{thm:laplace}}
\label{app:laplace}

From the \cref{lemma:dp_gcp}, we have that the Cox point synthesizer $\M_{\rm Lap}$ is $(\epsilon,0)$-DP if its intensity function $\lambda_{\GAMMA}$ is implicitly dependent on the original dataset $D$.

Consider a function $f(D)$ that takes an input point pattern dataset $D$ and outputs a vector $\GAMMA^0_D$, where $\GAMMA^0_D = (\gamma_{D,1}^0, \ldots, \gamma_{D,N}^0)$ and 
\begin{equation*}
    \gamma^0_{D,i} = \sum_{x \in D} \psi_i(x).
\end{equation*}
The $L_1$ sensitivity of this function is defined as
\begin{equation*}
    \Delta_1(f) = \max_{D \sim D'} \Vert f(D) - f(D') \Vert_1,
\end{equation*}
where $D \sim D'$ denotes datasets $D$ and $D'$ differing by a single point. For $f(D)$, the sensitivity simplifies to
\begin{equation*}
    \Delta_1(f) = \max_{p \neq q} \left| \frac{1}{|S_p|} + \frac{1}{|S_q|} \right|,
\end{equation*}
since the vector $f(D) - f(D')$ has at most two nonzero values, $\frac{1}{|S_p|}$ and $-\frac{1}{|S_q|}$, if the differing point $x$ in $D$ is located in the $p$-th cell and in the $q$-th cell in $D'$, respectively. Put $\GAMMA_D^1=(\gamma^1_{D,1},\ldots, \gamma^1_{D,N})$ as follows
\begin{equation*}
    \gamma^1_{D,i} = \gamma^0_{D,i} + {\rm Lap}\left(\frac{\Delta_1(f)}{\epsilon}\right),
\end{equation*}
where ${\rm Lap}(\theta)$ denotes i.i.d. Laplace-distributed random variables. Let $p(\TAU|D)$ be the Laplace density of the random vector $\GAMMA^1_D$. 
The ratio of two densities becomes
\begin{align*}
    \frac{p(\TAU|D)}{p(\TAU|D')} &=  \frac{\prod_{i=1}^N p(\tau_i|D)}{\prod_{i=1}^N 
 p(\tau_i|D')} =
    \frac{\prod_{i=1}^N \frac{\epsilon}{2\Delta_1(f)}\exp(-\frac{\epsilon}{\Delta_1(f)}|\tau_i -\gamma_{D,i}^0|)}{\prod_{i=1}^N \frac{\epsilon}{2\Delta_1(f)}\exp(-\frac{\epsilon}{\Delta_1(f)}|\tau_i- \gamma_{D',i}^0|)}\\
    &= \prod_{i=1}^N \exp\left(\frac{\epsilon}{\Delta_1(f)}\left(|\tau_i-\gamma_{D',i}^0| - |\tau_i-\gamma_{D,i}^0|\right)\right) \\
    &\le \prod_{i=1}^N \exp\left(\frac{\epsilon}{\Delta_1(f)}|\gamma_{D',i}^0-\gamma_{D,i}^0|\right) \quad (\text{triangle inequality}) \\
    &=\exp\left(\frac{\epsilon}{\max_{p \neq q} \left| \frac{1}{|S_p|} + \frac{1}{|S_q|} \right|}\sum_{i=1}^N |\gamma_{D',i}^0-\gamma_{D,i}^0|\right).
\end{align*}
If the single differing point in $x\in D$ is located in $k$-th cell and $x'\in D'$ is in $l$-th cell, then:
\begin{equation*}
    \frac{p(\TAU|D)}{p(\TAU|D')} \le \exp\left(\frac{\epsilon}{\max_{p \neq q} \left| \frac{1}{|S_p|} + \frac{1}{|S_q|} \right|} \left(\left|\frac{1}{|S_k|} + \frac{1}{|S_l|}\right| \right)\right) \le e^\epsilon.
\end{equation*}
Also, from the post-processing immunity (Proposition 2.1 in \citet{dworkAlgorithmicFoundationsDifferential2013}), releasing $\GAMMA = \max(0, \GAMMA^1_D)$ is also $(\epsilon,0)$-DP. Thus, from the \cref{lemma:dp_gcp}, the Cox point synthesizer $\M_{\rm Lap}$ using the piecewise constant intensity function in the \cref{thm:laplace} is $(\epsilon,0)$-DP. \qed

\section{Spatial point processes}
\label{app:pointprocess}

\subsection{Poisson point processes and Cox processes}
For the detailed properties of the Poisson and Cox processes, see \citet{mollerStatisticalInferenceSimulation2003}. 

\begin{definition}[Poisson point process]
    A spatial point process $\X$ on $S$ is a Poisson point process with intensity function $\lambda$ if the following properties hold.
    \begin{enumerate}
        \item For any $B\subseteq S$ with $\mu (B)<\infty$, the number of points $N (B)$ is Poisson random variable with mean $\mu (B)$ where $\mu$ is called as the \textit{intensity measure} of $\X$.
        \item For any $n\in \N$ and $B\subseteq S$ with $0<\mu (B)<\infty$, given $N (B)=n$, the points in $B$ are i.i.d. distributed with density $f (s)=\lambda (s)/\mu (B)$ where $\lambda(s)$ is the \textit{intensity function} of $\X$.
    \end{enumerate}
\end{definition}

\begin{definition}[log-Gaussian Cox process]
    A point process $\X$ is a log-Gaussian Cox process if its conditional distribution, given random intensity process $\lambda (\cdot)=\exp (Y (s))$, is a Poisson point process with intensity function $\lambda (\cdot)$, where $Y (s)$ is a Gaussian process. We refer to $Y (s)$ as \textit{latent Gaussian process}.
\end{definition}

\subsection{Spatial point processes on the linear network}
\label{app:linear_network}

We consider a special case where the spatial point processes are defined on the linear network $S=L$. Linear networks are different from usual graph structures in that the edges correspond to line segments on the plane. 

\begin{definition}[Linear network]
A line segment in the plane with endpoints $\mathbf{u},\mathbf{v}\in \R^2$ is a set of points $[\mathbf{u},\mathbf{v}] = \{t\mathbf{u} + (1-t)\mathbf{v} : 0\le t \le 1\}$. The endpoints $\mathbf{u}$ and $\mathbf{v}$ are called the \textit{nodes} of the line segment and the degree of a node is the number of line segments that meet at the node. A \textit{linear network} $L=\cup_i L_i$ is a finite collection of line segments $L_i$ such that the nodes of the line segments are connected to form a connected graph.
\end{definition} 

To utilize the log-Gaussian Cox process model on the linear network, we need to define a suitable covariance function for the latent Gaussian field. \citet{anderesIsotropicCovarianceFunctions2020} proposed an isotropic covariance function on the graph with Euclidean edges, which can be applied to the case of the linear network.

To begin with, consider the linear network $L=\cup_{i=1}^N L_i$ as a graph $G=(V,E)$ where $V \subset L$ is the set of nodes and $E$ is the set of edges. For the nodes $u,v\in V$, we define the shortest path distance $d_G(u,v)$ as the length of the shortest path between $u$ and $v$. Then, we define so-called \textit{conductance function} as
\begin{equation*}
    c(u,v) =\begin{cases}
        1/d_G(u,v) & \text{if } (u,v)\in E \\
        0 & \text{otherwise}.
    \end{cases}
\end{equation*}
Also, we define a matrix $|V|\times |V|$ matrix $C$ with
\begin{equation*}
    C(u,v) = \begin{cases}
        1+c(u_o) & \text{if } u=v=u_o \\
        c(u) & \text{if } u=v\neq u_o \\
        -c(u,v) & \text{otherwise},
    \end{cases}
\end{equation*}
where $c(u) = \sum_{v\in V}c(u,v)$ and $u_o$ is an arbitrary node in $V$. It can be shown that the matrix $C$ is a symmetric positive definite.

To define the resistance metric, we need two types of random fields $Z_\mu, Z_e$ over the linear network $L$. First, we define the random field $Z_\mu$ on the vertices $V$ as
\begin{equation*}
    (Z_\mu(v_1),\ldots,Z_\mu(v_n)) \sim \mathcal{N}(0,C^{-1}).
\end{equation*}
Then, for the arbitrary point $u\in L$, we define
\begin{equation*}
    Z_\mu(u) = (1-d(u))Z_\mu(\underline{u}) + d(u)Z_\mu(\overline{u})
\end{equation*}
where $d(u)=d_G(u,\underline{u})/d_G(\underline{u},\overline{u})$ if $u\notin V$ and $d(u)=0$ if $u\in V$. Here $\underline{u}$ and $\overline{u}$ denote the starting point and the endpoint of the edge that contains $u$, respectively.

Also, we define the Brownian bridge $B_e$ on the edge $e\in E$ so that $B_e(\underline{e})=B_e(\overline{e})=0$. Then, we define a random field $Z_e$ on the edge $e$ as
\begin{equation*}
    Z_e(u) = \begin{cases}
        B_e(\phi_e(u)) & \text{if } u\in e \\
        0 & \text{otherwise}
    \end{cases}
\end{equation*}
where $\phi_e(u)$ is a bijection from $e$ to $[\underline{e},\overline{e}]\subset \R$. The covariance function of $Z_e$ is given by
\begin{equation*}
    \cov(Z_e(u),Z_e(v)) = \begin{cases}
        \left\{\min(d(u), d(v))-d(u)d(v)\right\}d_G(\underline{u},\overline{u}) & \text{if } u,v\in e \\
        0 & \text{otherwise}.
    \end{cases}
\end{equation*}

Now we define the \textit{resistance metric} on the linear network.
\begin{definition}
    The resistance metric $d_R(u,v)$ between two points $u,v$ on the linear network $L$ is defined as
    \begin{equation}
        d_R(u,v) = \var(Z_L(u)-Z_L(v)), \quad u,v\in L
    \end{equation}
    where $Z_L$ is defined as
    \begin{equation}
        Z_L(u) = Z_\mu(u) + \sum_{e\in E}Z_e(u), \quad u\in L.
    \end{equation}
\end{definition}

The resistance metric $d_R(u,v)$ forms a metric on the linear network $L$ and can be used to define the isotropic covariance function on the linear network. For the proof, see Theorem 1 in \citet{anderesIsotropicCovarianceFunctions2020}.

\begin{theorem}[Theorem 1 in \citet{anderesIsotropicCovarianceFunctions2020}]
    Let $r_0:[0,\infty) \to \R$ be continuous, infinitely differentiable on $(0,\infty)$ and $(-1)^jr_0^{(j)}(t) \ge 0$ for all $t>0$ and $j\in\N \cup \{0\}$. Then, the function $r_0(d_R(u,v))$ is strictly positive definite over $L\times L$. Moreover, if $L$ is a 1-sum of trees and loops, then $r_0(d_G(u,v))$ is strictly positive definite over $L\times L$.
\end{theorem}

Thus, we can use the function $r_0$ as a correlation function for the log-Gaussian process on the linear network. Since the 1-sum structures which involve combinations of trees and loops, are not typical in the real-world data (e.g. road networks), we make use of the resistance metric $d_R$ in the following sections. Some parametric classes of $r_0$ are provided in \cref{tab:correlation}, where $K_\alpha$ denotes the modified Bessel function of the second kind.

\begin{table}
    \caption{Parametric classes of the correlation function $r_0$}
    \centering  
    \begin{tabular}{ccc}
        \toprule
        \textbf{Class} & \textbf{Formula} & \textbf{Parameters} \\
        \midrule
        Exponential & $r_0(t) = \exp\left(-\frac{t^\alpha}{\phi}\right) $ & $\alpha\in(0,1], \phi>0$ \\
        Matérn & $r_0(t) = \frac{2^{1-\alpha}}{\Gamma(\alpha)}\left(\frac{\beta}{\phi}\right)^\alpha K_\alpha\left(\frac{\beta}{\phi}\right)$ & $\alpha\in(0,\frac{1}{2}], \beta>0$ \\
        \bottomrule
    \end{tabular}
    \label{tab:correlation}
\end{table}

\subsection{K-function on a linear network}
\label{sec:kfunctionlinnet}

The K-function introduced by \citet{ripleySecondorderAnalysisStationary1976} has been extended to the linear network by \citet{okabeKFunctionMethodNetwork2001}. Suppose we have a linear network $L$ with the shortest path distance metric $d_L$ on the network $L$. The K-function on the network is defined as
\begin{equation}
    \hat K_{\rm net}(r) = \frac{|L|}{n(n-1)}\sum_{i=1}^{n}\sum_{j\ne i} \mathbf{1}(d_L(x_i,x_j)\le r).
\end{equation}
However, $\hat K_{\rm net}(r)$ is not a valid estimator of the K-function on the network because it depends on the geometric properties of the network. To address this issue, \citet{angGeometricallyCorrectedSecond2012} proposed the geometrically corrected K-function on the network as
\begin{equation}
    \hat K_{L}(r) = \frac{|L|}{n(n-1)}\sum_{i=1}^{n}\sum_{j\ne i} \frac{\mathbf{1}(d_L(x_i,x_j)\le r)}{m(x_i, d_L(x_i,x_j))},
\end{equation}
where the $m$ function is the correction factor defined as the number of perimeter points:
\begin{equation}
    m(u, r) = \# \{v \in L : d_L(u,v) = r\}.
\end{equation}
The inhomogeneous K-function on the network is also defined as 
\begin{equation}
    \hat K_{L}(r) = \frac{1}{\sum_{i=1}^{n}1/\lambda(x_i)}\sum_{i=1}^{n}\sum_{j\ne i} \frac{\mathbf{1}(d_L(x_i,x_j)\le r)}{\lambda(x_i)\lambda(x_j)m(x_i, d_L(x_i,x_j))}.
\end{equation}
Computational strategies for the K-function and the correction factor are provided by \citet{rakshitEfficientCodeSecond2019}.

\section{The effect of the tessellation type}
\label{sec:tessellation}

\cref{tab:simulation_square_triangular} shows the effect of the tessellation on the utility of the Poisson point synthesizer using the log-Gaussian Cox process ($\M_{\rm LGCP}$). The results are obtained from the simulation study with intensity function $\lambda_2$ to $\lambda_4$ and the privacy budget $\epsilon=0.1, 1.0, 10.0$. We compared two tessellations: the regular square tessellation (Sqr) and the triangular tessellation (Tri). For the square tessellation, we used the regular square grid with the same side length as the triangular tessellation. Also, like the triangular basis function, the square basis function is defined as the linear interpolation of the four nearest grid points. The results show that there is no significant difference in the utility of the Poisson point synthesizer between the two tessellations.

\begin{table}[ht!]
\centering
\caption{Simulation results for varying tessellation. 30 synthetic datasets were generated for each $\epsilon$.}
\label{tab:simulation_square_triangular}
\small
\begin{tabular}{cccccccc}
\toprule
$\lambda_i$ & $\epsilon$ & Method & pMSE (std) & npoints (std) & MISE \\ 
\midrule
\multirow{9}{*}{$\lambda_{2}$} & \multirow{3}{*}{0.1} & Ori & - & 77.6 (9.55) & - \\
 &  & Tri & 0.076 (0.005) & 77.2 (12.81) & 0.865 \\
 &  & Sqr & 0.077 (0.006) & 79.9 (16.43) & 1.080 \\
\cmidrule{2-6}
 & \multirow{3}{*}{1.0} & Ori & - & 77.6 (9.55) & - \\
 &  & Tri & 0.076 (0.005) & 77.2 (13.13) & 0.882 \\
 &  & Sqr & 0.076 (0.006) & 78.4 (16.28) & 0.814 \\
\cmidrule{2-6}
 & \multirow{3}{*}{10.0} & Ori & - & 77.6 (9.55) & - \\
 &  & Tri & 0.076 (0.005) & 76.9 (12.6) & 0.929 \\
 &  & Sqr & 0.077 (0.005) & 76.5 (16.21) & 0.643 \\
\cmidrule{1-6}
\multirow{9}{*}{$\lambda_{3}$} & \multirow{3}{*}{0.1} & Ori & - & 128.5 (12.34) & - \\
 &  & Tri & 0.052 (0.001) & 129.3 (15.85) & 0.114 \\
 &  & Sqr & 0.053 (0.002) & 132.7 (18.69) & 0.142 \\
\cmidrule{2-6}
 & \multirow{3}{*}{1.0} & Ori & - & 128.5 (12.34) & - \\
 &  & Tri & 0.052 (0.001) & 128.9 (16.86) & 0.108 \\
 &  & Sqr & 0.053 (0.002) & 131.1 (19.42) & 0.124 \\
\cmidrule{2-6}
 & \multirow{3}{*}{10.0} & Ori & - & 128.5 (12.34) & - \\
 &  & Tri & 0.052 (0.002) & 129.9 (18.76) & 0.110 \\
 &  & Sqr & 0.053 (0.002) & 130.7 (21.11) & 0.096 \\
\cmidrule{1-6}
\multirow{9}{*}{$\lambda_{4}$} & \multirow{3}{*}{0.1} & Ori & - & 57.3 (6.82) & - \\
 &  & Tri & 0.13 (0.009) & 56.7 (9.73) & 4.258 \\
 &  & Sqr & 0.128 (0.009) & 57.7 (12.87) & 4.505 \\
\cmidrule{2-6}
 & \multirow{3}{*}{1.0} & Ori & - & 57.3 (6.82) & - \\
 &  & Tri & 0.13 (0.009) & 58.1 (10.51) & 5.184 \\
 &  & Sqr & 0.129 (0.007) & 59.3 (12.1) & 5.077 \\
\cmidrule{2-6}
 & \multirow{3}{*}{10.0} & Ori & - & 57.3 (6.82) & - \\
 &  & Tri & 0.128 (0.008) & 57.6 (10.19) & 4.851 \\
 &  & Sqr & 0.13 (0.01) & 58.5 (12.6) & 4.670 \\
\bottomrule
\end{tabular}
\end{table}